\documentclass[twocolumn,preprintnumbers,amsmath,amssymb]{revtex4}
\usepackage{graphicx}

\begin{document}

%
%
\title{
Dynamic effects in double graphene-layer structures with inter-layer
resonant-tunneling negative conductivity 
}
\author{V~Ryzhii$^{1,3}$,
A~Satou$^{1,3}$,
T~Otsuji$^{1,3}$, M~Ryzhii$^{2,3}$,
V Mitin$^4$ 
and  M S Shur$^5$
}
\affiliation{
$^1$Research Institute for Electrical Communication,Tohoku University, Sendai 980-8577, Japan\\
$^2$ Department of Computer Science and Engineering,  University of Aizu, Aizu-Wakamatsu  965-8580, Japan\\
$^3$Japan Science and Technology Agency, CREST, Tokyo 107-0075, Japan\\
$^4$ Department of Electrical Engineering, University at Buffalo, Buffalo, New York 1460-1920, USA\\
$^{5}$Department of   Electrical, Electronics, and Systems Engineering and Physics, Applied Physics, and Astronomy, Rensselaer Polytechnic Institute, Troy, New York 12180, USA.\\
}

\begin{abstract}
We study  the dynamic effects 
in the double graphene-layer (GL)  structures with
 the resonant-tunneling (RT) and  the negative differential  
inter-GL conductivity. Using the developed model, which accounts for the excitation of self-consistent 
oscillations of the electron and hole densities and the ac electric field between GLs (plasma oscillations), 
 we calculate the admittance of
the double-GL RT  structures as a function of the signal  frequency 
and applied voltages, and the spectrum and increment/decrement of plasma oscillations.
Our results show that the electron-hole plasma in the  double-GL RT structures with realistic parameters is stable with respect to the self-excitation of plasma oscillations and aperiodic perturbations.
The stability of the  electron-hole plasma at the bias voltages corresponding to the  inter-GL  RT
and strong nonlinearity of the RT  current-voltage characteristics enable 
using  the  double-GL  RT  structures for detection of teraherz (THz)  radiation.
The excitation of plasma oscillations by the incoming THz radiation can result in a sharp resonant dependence of detector responsivity on radiation frequency and the bias voltage.
Due to a strong nonlinearity of the current-voltage characteristics of the double-GL   structures at RT  and the resonant excitation of plasma oscillations, the maximum responsivity, $R_V^{max}$, can  markedly exceed the values $(10^4 - 10^5)$~V/W at 
room temperature.
\end{abstract}

\maketitle
\newpage
\section{Introduction}

As demonstrated recently~\cite{1,2,3,4}, double graphene-layer (GL) structures with the inter-GL layers
forming relatively narrow and low energy
barriers for electrons or holes can be effectively used in novel  devices. The 
excitation of plasma oscillations in these structures, i.e., the excitation of spatio-temporal variations 
of the electron and hole densities in GLs and the spatio-temporal variations of the self-consistent electric 
field between GLs, by incoming terahertz (THz) radiation,  modulated optical radiation, or ultra-short 
optical pulses provides additional functional opportunities. In particular, the double-GL
structures with the inter-GL tunneling or thermionic conductance can be used in the resonant  
THz detectors and photomixers~\cite{5,6}. As was recently discussed~\cite{7,8,9} and realized experimentally~\cite{10}, the inter-GL  resonant tunneling (RT)  in 
  the double-GL structures, 
with the band diagrams  properly aligned by the applied voltage,  leads to inter-GL  negative differential
conductivity (NDC) and enables novel transistor designs with   the multi-valued current-voltage characteristics.  A  strong nonlinearity of the inter-GL current-voltage characteristics at the voltages near tunneling resonance can be used in double-GL-based 
  frequency multipliers~\cite{8}, detectors~\cite{5}, and other microwave and THz devices.
However, NDC might, in principle, result in the instability of stationary states, modifying or even  harming normal operation of double-GL transistors and two-terminal devices.

In this paper, we consider the double-GL structures with tunneling transparent barrier layers exhibiting
RT and NDC. Using the proposed model, we calculate the device admittance and  demonstrate that   
a steady-state  current flow in these structures is stable with respect to the self-excitation
 of plasma oscillations  and aperiodic perturbations despite NDC. The  incoming electromagnetic radiation, particularly, 
 THz radiation can result in an effective resonant excitation of plasma oscillations,  which can be used for
 the THz detection.
  We calculate the rectified current and the responsivity
of the THz detectors based on the  double-GL  RT structures and
 show, that due to a strong nonlinearity of the double-GL current-voltage characteristics and the possibility
 of the resonant excitation of plasma oscillations by incoming THz radiation, the structures under consideration can
 serve as THz detectors exhibiting very high responsivity.

\begin{figure*}[t]\label{Fig.1}
\vspace*{-0.4cm}
\begin{center}
\includegraphics[width=7.5cm]{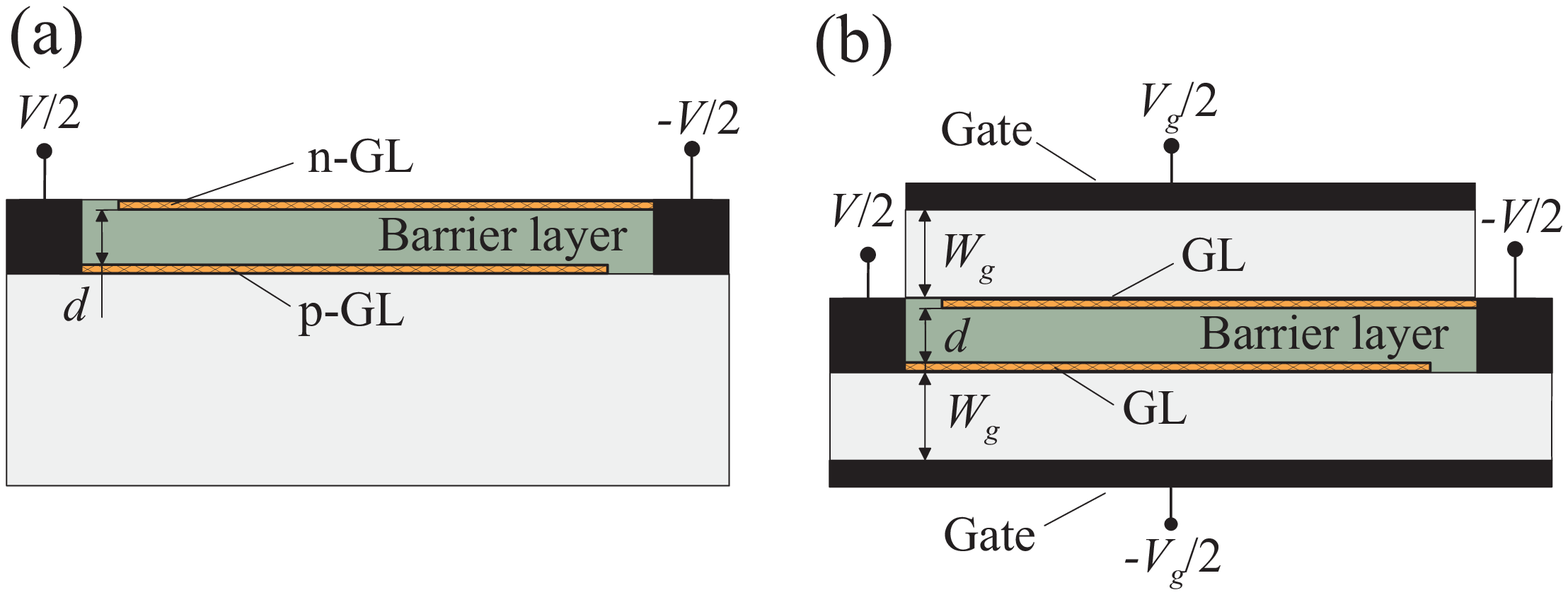}\hspace*{+1.0cm}
\includegraphics[width=7.5cm]{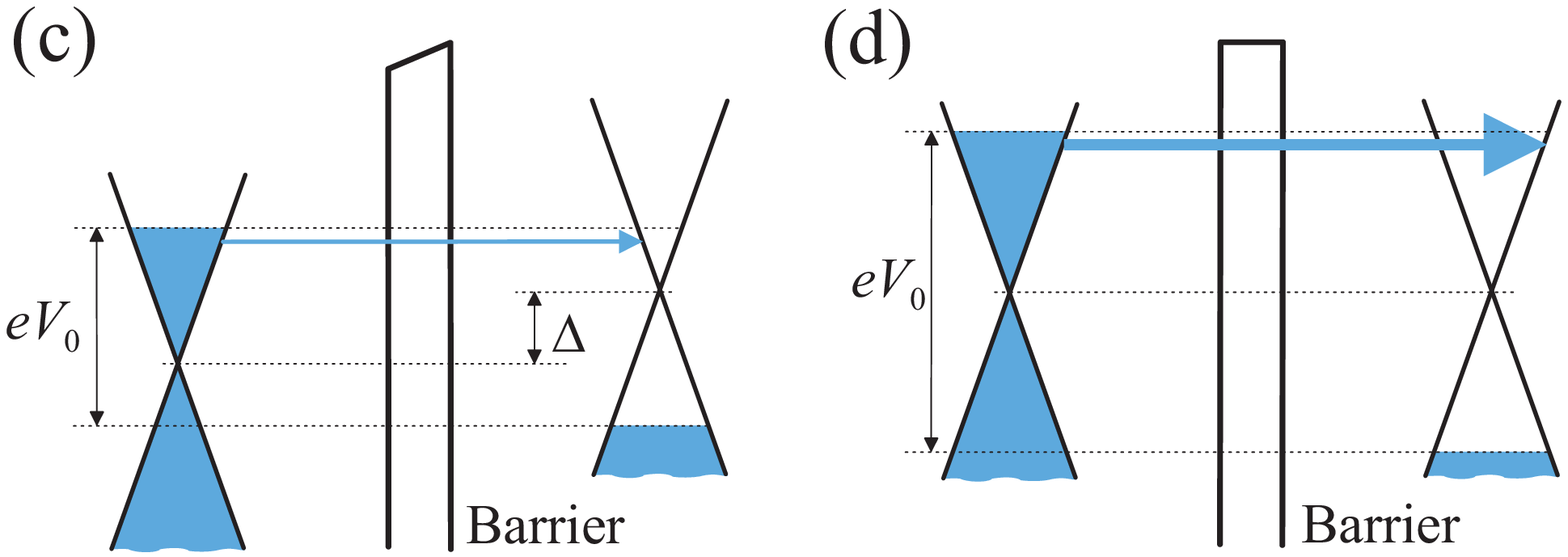}
\caption{
Schematic views of (a) a doped double-GL structure and (b) a gated double-GL structure
with "electrical" doping, (c) band diagram at $V_0 < V_{t}$ and (b) at $V_0 = V_{t}$.
Horizontal arrows correspond to inter-GL non-resonant and RT transitions }
\end{center} 
\end{figure*}

\section{Model  equations}

We consider the double-GL structures either
with  chemically doped GLs  (figure 1(a)) or with  undoped GLs but sandwiched between
two highly conducting gates (figure 1(b)). In the first case,  one of the GLs is doped by donors, whereas the  other one is doped by acceptors, so that the electron and hole sheet  densities in the pertinent GLs are equal
tothe dopant density (donors and acceptors) $\Sigma_i$. 
In the second case, the voltage $V_g$ applied between the gates, induces the electrons and holes with the sheet charge densities $\pm e\Sigma_i \propto V_g/W_g$, where $W_g$ is the thickness of the layers separating GLs and the gates (see figure 1(b)). 

 Each  GL is supplied
by an Ohmic side contact (left-side contact to lower GL and right-side contact to upper GL) between which
the bias voltage $V_0$ is applied. This voltage  affects the electron and hole steady-state densities
 $\Sigma_{0}$:

\begin{equation}\label{eq1}
\Sigma_{0}=  \Sigma_i + \frac{\kappa\,V_0}{4\pi\,ed}.
\end{equation} 
 Here $e = |e|$ is the value of the electron charge, $\kappa$ is the dielectric constant, and $d$ is the spacing between GL (i.e., the thickness of the inter-GL barrier). 
Figures 1(c) and 1(d) show the  band diagrams of the double-GL structures under consideration 
at $V_0 < V_{t}$ and $V_0 = V_{t}$, respectively.
The energy difference, $\Delta$, between the Dirac points in GLs is determined 
by the following equation (see figure 1(c)):

\begin{equation}\label{eq2}
\Delta =  2\varepsilon_F - eV_{0},
\end{equation} 
where $\varepsilon_F$ is the Fermi energy of electrons and holes in the pertinent GL.
In practically interesting cases, the electron and hole systems in GLs are degenerated even at room temperatures. This occurs when $\Sigma_i$ is sufficiently large in comparison with the equilibrium density
$\Sigma_T$
: $\Sigma_i \gg \Sigma_T = \pi k_B^2T^2/6\hbar^2v_W^2$, where $\hbar$ and $k_B$ are the Planck 
and Boltzmann constants, respectively, $v_W \simeq 10^8$~cm/s is the characteristic velocity of electrons and holes in GLs, and $T$ is the temperature.
In such a case,

\begin{equation}\label{eq3}
\varepsilon_F = \hbar\,v_W\sqrt{\pi\biggl(\Sigma_i + \frac{\kappa\,V_0}{4\pi\,ed}\biggr)},
\end{equation} 
 Using equations (1) and (2) and  the alignment condition $\Delta = 0$, we obtain the following formula for the alignment voltage:

$$
eV_{t}=  \frac{\kappa\hbar^2v_W^2}{2\,e^2d}\biggl(1 +\sqrt{1 + \frac{16\pi\,e^4d^2\Sigma_i}
{\kappa^2\hbar^2v_W^2}}\biggr) 
$$
\begin{equation}\label{eq4}
\gtrsim 2\hbar\,v_W\sqrt{\pi\Sigma_i} = 2\varepsilon_{Fi}.
\end{equation}
A  difference between the quantities $eV_{t}$ and $2\varepsilon_{Fi}$ is due to the quantum capacitance effect~\cite{11}.

The local value of the  inter-GL resonant-tunneling current density as a function of the bias voltage $V_0$ is given by ~\cite{7,8}
\begin{equation}\label{eq5}
j_t = j_t^{max} \exp\biggl[- \biggl(\frac{V_0 - V_{t}}{\Delta V_{t}}\biggr)^2\biggr],
\end{equation}
where $ j_t^{max} $ is the peak value of the current density and $\Delta V_t = 2\sqrt{2\pi}\hbar\,v_W/el$ determines the peak width,  $l$ is the coherence length (the characteristic size of the  ordered areas in GLs).

Using equation (5) , at small signal  variations of the local potential difference  between GLs  $(\delta\varphi_{+} - \delta\varphi_{-})$ , the variation of the RT  current density can be presented as 

\begin{equation}\label{eq6}
\delta j_t \simeq \sigma_t(\delta\varphi_{+} - \delta\varphi_{-}) + \beta_t(\delta\varphi_{+} - \delta\varphi_{-})^2.
\end{equation}
Here
$\sigma_t =  (dj_t/dV)|_{V =V_0}$ 
is the differential tunneling conductivity and
$\beta_t = \displaystyle \frac{1}{2}(d^2 j_t/d V^2)|_{V = V_0}$ determines the nonlinearity of the RT current-voltage characteristics:
 
\begin{equation}\label{eq7}
\sigma_t = - \frac{2j_t^{max}}{\Delta V_t}\biggl(\frac{V_0 - V_t}{\Delta V_t}\biggr)
\exp\biggl[- \biggl(\frac{V_0 - V_t}{\Delta V_{t}}\biggr)^2\biggr],
\end{equation}

\begin{equation}\label{eq8}
\beta_t = \frac{2j_t^{max}}{(\Delta V_t)^2}\biggl[2\biggl(\frac{V_0 - V_t}{\Delta V_t}\biggr)^2 - 1\biggr] \exp\biggl[- \biggl(\frac{V_0 - V_t}{\Delta V_{t}}\biggr)^2\biggr].
\end{equation}
As follows from equation (7), $\sigma_t$ changes its sign at $V_0 = V_{t}$ and becomes negative when $V_0 > V_{t}$. The maximum value of $|\sigma_t|$, which is achieved at
$V_0 - V_{t} = \pm \Delta V_{t}/\sqrt{2}$, is equal to 
$|\sigma_t| = \sqrt{2}e^{-1/2}(j_t^{max}/\Delta V_t)$.  Equation (8)  for the value $\beta_t$ at
$V_0 = V_{t}$  yields $\beta = - 2j_t^{max}/(\Delta V_{t})^2$.

The spatial 
distributions of $\delta\varphi_{+}(x)$ and $\delta\varphi_{-}(x)$ in the  GL plane (along the axis $x$)  can be found from 
linearized  hydrodynamic equations (adopted for  the energy spectra of the electrons and holes in GLs~\cite{12}) coupled with the Poisson equation in the gradual channel approximation~\cite{13}. 
We limit our treatment to the in-phase perturbations of the electron and hole densities.
For such perturbations, the self-consistent ac electric field $\delta E =(\delta\varphi_+ - \delta\varphi_-)/d)$  is located between GLs.
The difference in the local values of the ac potentials of the upper and lower GLs causes the inter-GL current, which can either
increase or decrease with varying $(\delta\varphi_+ - \delta\varphi_-)$ depending on the value of
$V_0 - V_t$.
For the out-off-phase perturbations of the electron and hole densities, the ac electric field
is located mainly outside the double-GL structure, so that the inter-GL tunneling is insignificant and no effects associated with NDC can be expected.

 Calculating the ac potentials, one can neglect the nonlinear component
of the ac inter-GL current. i.e., the second term in the right-hand side of equation~(6)
(see, however, Sec.~IV) and searching  for the ac potential in the following form:  
$\delta\varphi_{\pm} = \delta\varphi_{\pm}(x)\exp(-i\omega t)$,
where $\omega$ is the complex signal frequency. In this case,
the  system of equations in question  can be reduced 
to the equations for the ac components of the potential at  the frequency $\omega$~\cite{3,4}:

\begin{equation}\label{eq9}
\frac{d^2\delta\varphi_{+}}{dx^2} 
+\frac{\omega(\omega + i\nu)}{s^2} (\delta\varphi_{+} - \delta\varphi_{-})=  - i\delta j(\omega + i\nu)\frac{4\pi\,d}{\kappa\,s^2}, 
\end{equation}
\begin{equation}\label{eq10}
\frac{d^2\delta\varphi_{-}}{dx^2} 
+\frac{\omega(\omega + i\nu)}{s^2} (\delta\varphi_{-} - \delta\varphi_{+})= i\delta j(\omega + i\nu)\frac{4\pi\,d}{\kappa\,s^2}. 
\end{equation}
Here $\nu$
is the collision frequency of electrons and holes in GLs with impurities and 
acoustic phonons and $s$
is the characteristic velocity of plasma waves in double-GL structures. 
Since electrons and holes belong to different GLs separated by a relatively
high barrier, we have disregarded the  electron-hole scattering and, hence, the effect of mutual electron-hole drag~\cite{12}.
The characteristic velocity $s$ in the double-GL structures (similar to that in the two-dimensional
electron or hole channels in the standard semiconductors with  metal gates)
is determined by the net dc electron and hole densities 
$\Sigma_0$ and the inter-GL layer thickness $d$~\cite{14,15,16,17}. In double-GL structures with the degenerate electron-hole plasma, $s =\sqrt{4\pi\,e^2\Sigma_0 d/\kappa\,m}$,
where $m \propto \sqrt{\Sigma_0} $ is the "fictitious" mass of electrons and holes in GLs. This implies that  
$s \propto \Sigma_0^{1/4}$~\cite{14}. 
The  value of $s$ in the GL structures under consideration can be fairly high, always exceeding 
the characteristic velocity of electrons and holes in GLs $v_W$~\cite{14,18}.
%
Considering equations~(6), (9), and (10),  we obtain

\begin{equation}\label{eq11}
\frac{d^2\delta\varphi_{+}}{dx^2} 
+\frac{(\omega + i\nu_t)(\omega + i\nu)}{s^2} (\delta\varphi_{+} - \delta\varphi_{-})= 0, 
\end{equation}
\begin{equation}\label{eq12}
\frac{d^2\delta\varphi_{-}}{dx^2} 
+ \frac{(\omega + i\nu_t)(\omega + i\nu)}{s^2}(\delta\varphi_{-} - \delta\varphi_{+})= 0. 
\end{equation}
Here  
$\nu_t = 4\pi\sigma_td/k$ is the characteristic frequency of the
 inter-GL tunneling. At $V_0 = V_{t}$,   $\sigma_t = 0$ and $\nu_t = 0$, while at $V_0 \gtrsim V_t$,  $\sigma_t <  0$ and, hence $\nu_t < 0$.

\begin{figure}[t]
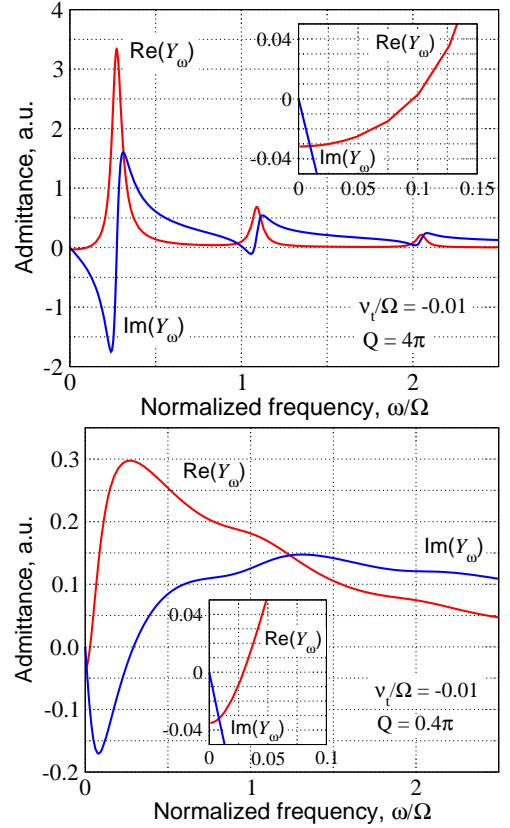
\label{Fig.2}
\vspace*{-0.4cm}
\begin{center}
\includegraphics[width=6.5cm]{DGRT_STAB_DET_F2a.eps}
\\
\includegraphics[width=6.5cm]{DGRT_STAB_DET_F2b.eps}
\caption{Real and imaginary parts, Re~$Y_{\omega}$ and Im~$Y_{\omega}$ of the admittance  versus 
signal frequency $\omega$ normalized by the plasma frequency $\Omega$
 for double-GL RT structures with different 
 plasma oscillation quality factor $Q = \Omega/\nu$ ($\nu_t/\Omega = -0.01$)
 Insets show the frequency dependences in the range of  low signal frequencies. 
 }
\end{center}
\end{figure}

Assuming that the total voltage between the contacts to GLs is equal to $V = V_0 + \delta V_{\omega}$, where $\delta V_{\omega}$ is the small signal voltage component, one can
 use the following boundary conditions for equations~(11) and (12):

\begin{equation}\label{eq13}
\delta\varphi_{\pm} |_{x = \pm L}  = \pm\frac{\delta V_{\omega}}{2}\exp(-i\omega\,t) ,\qquad \frac{d \delta\varphi_{\pm}}{dx} \biggl|_{x = \mp L}  = 0.
\end{equation} 
 The latter boundary conditions reflect the fact that the electron and hole 
 currents are equal to
 zero at the disconnected edges of GLs (at $x = -L$ in the upper GL and 
 at $x = L$ in the lower GL),  while the difference of the ac potentials $\delta\, V_{\omega}$ can generally be
 nonzero.

Solving equations~(11) and (12) with boundary conditions (13), we obtain

\begin{equation}\label{eq14}
\delta\varphi_{+}  -  \delta \varphi_{-} = \delta V_{\omega}\biggl(\frac{\displaystyle\frac{\cos\gamma_{\omega} x}
{\gamma_{\omega}\sin\gamma_{\omega} L}}
{\displaystyle\frac{\cos\gamma_{\omega} L}{\gamma_{\omega}\sin \gamma_{\omega} L} - L}\biggr),
\end{equation}
where $\gamma_{\omega} = \sqrt{2(\omega + i\nu_t)(\omega + i\nu)}/s$.

\section{Admittance of the double-GL resonant-tunneling structures}
 
\begin{figure*}[t]
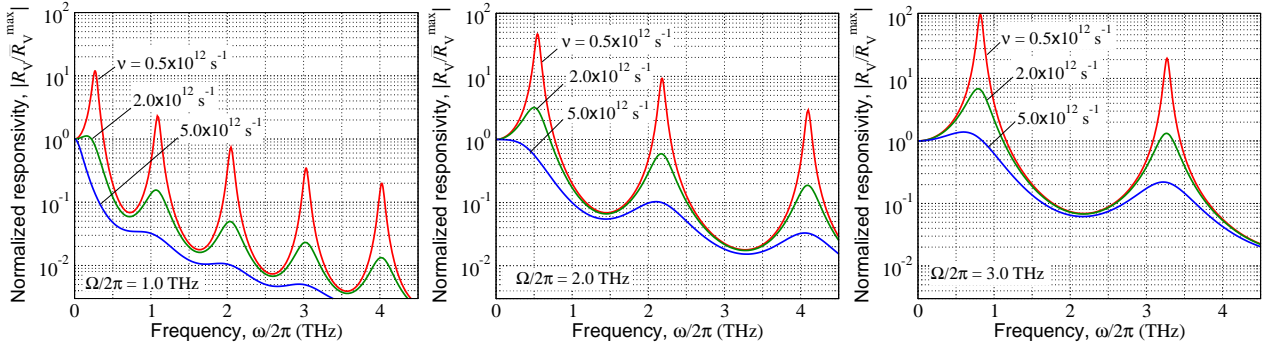
\label{Fig.3}
\vspace*{-0.4cm}
\begin{center}
\includegraphics[width=5.5cm]{DGRT_STAB_DET_F3a.eps}
\includegraphics[width=5.5cm]{DGRT_STAB_DET_F3b.eps}
\includegraphics[width=5.5cm]{DGRT_STAB_DET_F3c.eps}
\caption{Normalized responsivity $R_V/{\overline R_V}^{max}$ at versus signal frequency $\omega$
for different values of the electron and hole collision frequency $\nu$ and 
plasma  frequency $\Omega$.} 
\end{center}
\end{figure*}

First, we calculate the small-signal  admittance, of the double-GL RT structures,
$Y_{\omega} = \delta J_{\omega}/\delta V_{\omega}$, where 
\begin{equation}\label{eq15}
\delta J_{\omega} = H\biggl(-i\frac{\kappa\omega}{4\pi\,d} + \sigma_t\biggr)\int_{-L}^{L}dx (\delta\varphi_{+}  -  \delta \varphi_{-}) 
\end{equation}
is the net ac current, including the displacement current and  $H$ is the width of the double-GL structure in the direction 
perpendicular to the currents. Using equations (14) and (15), we find

$$
Y_{\omega} = -i\biggl(\frac{\kappa\,HL}{2\pi\,d}\biggr)\frac{(\omega + i\nu_t)}{[\gamma_{\omega}L(\cot \gamma_{\omega} - \gamma_{\omega}L)]}
$$
\begin{equation}\label{eq16}
=
-i\biggl(\frac{\kappa\,HL}{2\pi^2\,d}\biggr)\sqrt{\biggl(\frac{\omega + i\nu_t}{\omega + i\nu}\biggr)}
\frac{\Omega}{(\cot \gamma_{\omega} - \gamma_{\omega}L)}.
\end{equation}
Here we have introduce the plasma frequency $\Omega = (\pi\,s/\sqrt{2}L)$, so that 
$\gamma_{\omega}L = \pi\sqrt{(\omega + i\nu_t)(\omega + i\nu)}/\Omega$.
In the limit $\omega = 0$, equation (16) yields
$Y_0 = 2HL\sigma_t$ if $\Omega \gg \nu$. This implies that in this case the dc admittance
is determined by the inter-GL conductivity. 
When  $\Omega^2 \ll |\nu_t|\nu $, we find 
$Y_0 = (\kappa\,2HL/4\pi\,d)(\Omega^2/\pi\nu) \propto \sigma_0$, where  $\sigma_0 \propto (e^2\Sigma_0/m\nu)$ stands for the dc conductivity of GLs. 


Figure~2 shows the frequency dependences of the real and imaginary parts of the admittance  calculated
using equation (16) for different values of the plasma oscillation quality factor 
$\Omega/\nu$ ($Q = 4\pi$ and $Q = 0.4\pi$) and $\nu_t/\Omega  = - 0.01$.
As seen, Re~$(Y_{\omega})$ is negative in a narrow range of small frequencies.
This is due to NDC at $(V_0 - V_T)\gtrsim \Delta V_t$ associated with RT.
However, at higher signal frequencies, Re~$(Y_{\omega})$ is positive.
One can also see that Im~$(Y_{\omega})$ does not change its sign in the frequency region,
where Re~$(Y_{\omega}) < 0$. This implies that the electron-hole plasma is stable, at least
for  the chosen parameters.
Indeed, the stability of stationary state of the electron-hole plasma at given bias voltage $V_0$
is determined using the condition $Z_{\omega} = Y_{\omega}^{-1} = 0$.
Considering the latter, we obtain the following condition for the damped or growing 
of perturbations  of the electron and hole densities, i.e., the following dispersion equation for plasma oscillations: 

\begin{equation}\label{eq17}
\cot \gamma_{\omega}L - \gamma_{\omega}L =  0.
\end{equation} 
Equation ~(17) determines  the complex frequency $\omega = \omega^{\prime} + i\omega^{\prime\prime}$, 
where $\omega^{\prime}$ is real value of possible plasma oscillations and $\omega^{\prime\prime}$
is
 their damping/growth rate.
 For the structures with the characteristic plasma frequency $\Omega \gg \nu, |\nu_t|$, 
 equation (17) yields $\omega^{\prime} = \omega_n$,
$(n =0,1,2,3,..)$ and $\omega^{\prime\prime} = \Gamma$, where

\begin{equation}\label{eq18}
 \omega_0 \simeq \frac{0.86}{\pi}\Omega, \qquad \omega_n \simeq n\Omega + \frac{\Omega}{\pi^2n},\qquad
 \Gamma \simeq -\frac{\nu + \nu_t}{2}.
\end{equation}
 
If  $\Omega \ll \nu$,  we find $\omega^{\prime} = 0$ from equation~(17)

\begin{equation}\label{19}
 \Gamma \simeq -\frac{\Omega^2}{\pi^2\nu} - \nu_t.
\end{equation}

The plasma frequency can vary in a wide range depending  on the structure length $2L$.
In particular, setting $s = (2 - 4)\times 10^8$~cm/s and $2L = 1~\mu$m (as in Ref.~\cite{10}), one obtains $\Omega/2\pi \simeq 1.4 - 2.8$~THz (so that according to equation (18) $\omega_0 \simeq 0.554 - 1.1$~THz). But if $2L = 10~\mu$m, one can get 
 $\Omega/2\pi \simeq 0.14 - 0.28$~THz.

As follows from equations~(18) and (19), the plasma instability (the increment, i.e,the
 growth rate $\Gamma > 0$) is possible   if $\nu_t < - \nu$ in the structures with a high quality factor $Q = \Omega/\nu \gg 1$ or
$\nu_t < -  \Omega^2/\pi^2\nu$
in the structures with $Q \ll 1$. In the former case, plasma oscillations with the frequencies $\omega_n$
can self-excite, while in  latter case, the growth of the perturbations of the electron and hole densities is aperiodic, which could potentially result in the  domain formation.
In such  situations, the stationary current flow between GLs could be unstable. However, in real double-GL
structures the value of the differential  inter-GL  RT conductivity is not sufficiently large to provide the condition $\nu_t < - \nu$. 
To estimate the real value of $\nu_t$, we assume that  $j_t^{max} = (5 - 30)$~A/cm$^2$~\cite{8,9,10}
and 
$l = 100$~nm, so that $V_t \simeq 30$~mV. This yields $|\sigma_t| = (143 - 858)~$~S/cm$^2$. If $d = 4$~nm and $\kappa = 4$ (hBN four atomic layers thick barrier),
one obtains $|\nu_t| \simeq (1.6 - 9.7)\times10^8$~s$^{-1}.$
As seen, at realistic $\nu = (10^{11} - 10^{13})$~s$^{-1}$,  the value $|\nu_t| \ll \nu$.
This is in contrast 
to the double-barrier RT devices based on InGaAs-AlAs, where  the frequency $|\nu_t|$ can 
be rather high, being of the order of or even exceeding the electron collision frequency.
This is due to a very high peak current density and modest width of the tunneling resonance $\Delta V_t$ in the double-barrier RT diodes.
This can lead to the instability of the stationary current with respect to the self-excitation of plasma 
oscillations ~\cite{19}. For instance, in one of the best resonant-tunneling diode~\cite{20},  
$\kappa = 12$,$|\sigma_t| = 3.3\times10^6$~S/cm$^2$ 
and $d = 31$~nm, so that $\nu_t \simeq 10^{13}$~s${-1}$.

In relatively long double-GL structures, inequality $\nu_t < - \Omega^2/\pi^2\nu$ can be satisfied if  the
 collision frequency $\nu$ is large and, hence, the mobility of electrons and holes is low.
Indeed, setting $\Omega/2\pi = 0.14 - 0.28$~THz and $\nu = (1 - 5)\times10^{13}$~s$^{-1}$, the latter 
inequality needs
$|\nu_t| > (1.6 - 7,8)\times 10^9$~s$^{-1}$. The latter condition is not met for
the double-GL structures  considered recently~\cite{8,9,10}.

 \section{Resonant detection of  radiation}

The ac potential $\delta V_{\omega} $ between the contacts to GLs can arise 
due  incoming electromagnetic radiation received by an antenna.
This results in the excitation of plasma oscillations in the double-GL structure 
described by equation~(14). The ac  potential drop $(\delta\varphi_{+}  -  \delta \varphi_{-})$ causes 
not only the linear component of the tunneling current $\delta J_{\omega}$ but also the rectified dc current,
 $\delta J_0$,
associated with the nonlinear (quadratic) component. The rectified ac current is
given by the following formula:

\begin{equation}\label{eq20}
\delta J_{0} =  H\int_{-L}^{L}dx \beta_t |\delta\varphi_{+}  -  \delta \varphi_{-}|^2,
\end{equation}

 This rectified component of the inter-GL RT current can be used for detection of THz signals. 
Using equations~(14) and (20), we obtain the following formulae for the rectified current $\delta J_0$ and
the detector Volt-Watt responsivity $R_V$

\begin{equation}\label{eq21}
\delta J_0 = 
\frac{\beta_t(\delta V_{\omega})^2LH}
{|\cos\gamma_{\omega} L -\gamma_{\omega}L\sin \gamma_{\omega} L|^2}
\int_{-1}^{1}d\xi
|\cos(\gamma_{\omega}L\xi)|^2.
\end{equation}

\begin{equation}\label{eq22}
R_V = \frac{{\overline R_V}}{2}
\frac{\int_{-1}^{1}d\xi
|\cos(\gamma_{\omega}L\xi)|^2}
{|\cos\gamma_{\omega} L -\gamma_{\omega}L\sin \gamma_{\omega} L|^2}.
\end{equation}

Here the characteristic responsivity ${\overline R_V}$ is given by

\begin{equation}\label{eq23}
{\overline R_V} = \frac{2\pi}{cG}\biggl(\frac{\beta_t}{\sigma_t^0}\biggr) = \frac{2\pi}{cG}\biggl(\frac{\beta_tV_0}{j_t^0}\biggr),
\end{equation}
where $c$ is the speed of light in vacuum, $G \simeq 1.5$ is the antenna gain factor, $\sigma_t^0 = j_t^{0}/V_0$ and $j_T^{0}$ are the inter-GL dc conductivity and current density, respectively.
Considering that at $V_0 = V_t$ (when the resonant-tunneling current exhibits a maximum) above expression yields $\beta_t = j_t^{max}/(\Delta V_t)^2$, we arrive at the following formula

\begin{equation}\label{eq24}
{\overline R_V}^{max}= - \frac{2\pi}{cG}\biggl[\frac{V_0}{(\Delta V_t)^2}\biggr].
\end{equation}
Equation~(23) provides the  frequency dependence similar to the responsivity $R_V^{n}$ of the double-GL detectors
using the  nonlinearity of the inter-GL relatively smooth current-voltage characteristic with the tunneling assisted by electron scattering (non-resonant tunneling detector) considered by us recently~\cite{5}.
However, there are two  distinctions (apart from the difference in the $R_V$ sign). First, the absolute value of ${\overline R_V}$
is much larger than  ${\overline R_V^{n}}$. This is due to a significantly larger value of 
$|\beta_t|$
in the RT detectors associated with a high and sharp RT peak. Indeed, considering equation~(24)
and the pertinent equation in ~\cite{5}, one can arrive at
\begin{equation}\label{eq25}
\frac{{\overline R_V}^{max}}{{\overline R_V^{n}}} \simeq \biggl(\frac{V_0}{\Delta V_t}\biggr)^2.
\end{equation}
Setting $\Delta V_t = 30 - 90$~mV and $V_0 = 1000$~mV, we obtain ${\overline R_V}/
{\overline R_V^{n}}
\simeq 10^2 - 10^3$. At the above parameters, assuming that $j_t^0 = j_t^{max}$, 
one obtains ${\overline R_V} \simeq (1.5- 13.9)\times 10^4$~V/W.
Second, the responsivity peaks of the inter-GL RT detector width is characterized by the collision frequency $\nu$ (because $\nu_t = 0$),
but in the detectors using non-resonant tunneling the width of the peaks is determined by
$(\nu + \nu_t)$ with $\nu_t > 0$. 
The possibility to achieve very high values of the characteristic responsivity 
${\overline R_V}$ is connected with a large value of $\beta_t$ and, hence,
 small values of the width of tunneling resonance  $\Delta V_t$. One can assume that the latter quantity weakly depends on the temperature~\cite{8}, so that the responsivity can be very
 high even at room temperature.

\begin{figure}[t]\label{Fig.4}
\vspace*{-0.4cm}
\begin{center}
\includegraphics[width=6.0cm]{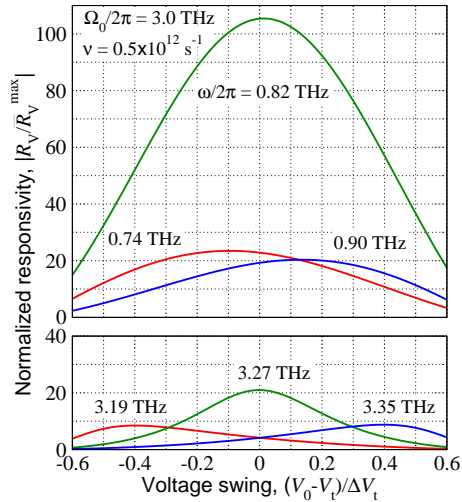}
\caption{Dependence of normalized responsivity $R_V/{\overline R_V}$ versus
 bias voltage swing $(V_0 - V_t)/\Delta V_t$.n  at different signal  frequency $\omega$
 near the zeroth plasma resonance (upper panel) and near the first plasma resonance
 (lower panel).}
\end{center}
\end{figure}

As follows from equation (22), the frequency dependence of the responsivity  exhibits sharp maxima
at the plasma resonant frequencies $\omega = \omega_n$, where the frequencies  $\omega_n$ are given
by equation~(16). The widths of the peaks are determined by the parameter $\Gamma \propto \nu$
(at the RT resonance $\nu_t = 0$). 
At the plasma resonances, $|R_V| \gg {\overline R_V}$. Thus, very high values of the responsivity
of  the detectors in question  can be achieved due to combining of the tunneling and plasma resonances.
  
Since the resonant plasma frequencies fall into the THz range, the detector under consideration
can be particularly useful for the resonant detection  of THz radiation. 

Figure~3 demonstrates examples of the frequency dependences of the responsivity calculated
using equation (22) for the double-GL RT structures with different values of the electron and hole collision frequency $\nu$ and different values of the plasma frequency $\Omega$.
The obtained dependences of the responsivity versus signal frequency exhibit
several resonant peaks associated with the plasma oscillations.
The highest peaks correspond to the  zeroth resonances at the frequency $\omega_0 < \Omega$,
while the other resonances  correspond to multiples of the plasma frequency $\Omega$   
(see equation ~(18)). One can see that $R_V > {\overline R_V}^{max}$ not only at the zeroth
plasma resonance, but also a higher resonances. The number of such resonances depends
on the quality factor $Q$.  The responsivity is very high
due RT with sharp maximum at the current voltage-characteristics 
even at the moderate quality factors. However, it is   much higher at the pronounced plasma resonances. 
 
The high values of the responsivity with the frequency characteristics of figure~3  
are associated with the combination of the tunneling and plasma resonances.
The deviation from the tunneling resonance leads to lowering of the responsivity.
If $V_0 \neq V_t$, the factor ${\overline R_V}$ in equation (22) becomes smaller than
${\overline R_V}^{max}$:
$$
{\overline R_V} = {\overline R_V}^{max}\biggl[2\biggl(\frac{V_0 - V_t}{\Delta V_t}\biggr)^2 - 1
\biggr]
$$
\begin{equation}\label{eq26}
\times\exp\biggl[-\biggl(\frac{V_0 - V_t}{\Delta V_t}\biggr)^2\biggr].
\end{equation}

Figure~4 shows the dependences of the normalized responsivity, $R_V/{\overline R_V}^{max}$ on  the voltage swing $(V_0 - V_t)/\Delta V_t$ calculated using equation (26)
for different signal frequencies in the vicinity of the zeroth and first plasma resonances.
The plasma frequency at $V_0 = V_t$  was chosen  to be $\Omega_0/2\pi = 3.0$~THz .
 The plasma frequency at different values of $V_0$
is given by $\Omega = \Omega_0[1 + (V_0 - V_t)/{\overline V_0}]^{1/4}$, where ${\overline V_0} = 4\pi\,ed/\kappa\Sigma_i$. 
At $\kappa = 4$, $d = 4$~nm, and $\Sigma_i = (1 - 5)\times 10^{12}$~cm$^{-2}$,
${\overline V_0} \simeq 180 - 900$~mV.
We set ${\overline V_0}/\Delta V_t = 6$ and $\nu = 0.5\times10^{12}$~s$^{-1}$.
As seen from figure~4, an increase in the absolute value of the voltage swing $|V_0 - V_t|/\Delta V_t$ results in a marked drop of the responsivity.
It is also seen that detuning of the plasma resonance leads to a significant decrease in the responsivity
 (compare the curves for
the plasma resonances at $\omega/2\pi \simeq 0.82$~THz and  $\omega/2\pi \simeq 3.27$~THz
with those  corresponding to a detuning $\delta \omega/2\pi = \pm 0.08$~THz.
Slightly different maximum values of the responsivity shifted  from   the plasma resonances 
are due to a small asymmetry of the resonant peaks.
Different positions of these maxima are associated with the dependence of the plasma frequency on $V_0$. 

The excitation of plasma oscillation by electromagnetic signals can be used not only for the resonant reinforcement  of the  rectified current  (i.e., the detector responsivity), but
also for a more effective generation of higher harmonics~\cite{8}.

\section{Conclusions}
In summary, we considered the dynamic behavior of the double-GL RT structures.
We calculated the frequency-dependent admittance and the responsivity of the double-GL
RT structures  to the incoming signals  as functions of the structural parameters, bias voltages, and frequency.
It was demonstrated that the stationary states of the electron-hole plasma are stable with respect to the self-excitation of plasma oscillations and aperiodic perturbations for the structures with realistic parameters.
As shown, the responsivity exhibits sharp resonant maxima corresponding to the excitation
of  plasma modes by incoming electromagnetic radiation. 
The plasma oscillations and the pertinent responsivity peaks are in the THz range.
The responsivity of the double-GL RT detectors  operating at room temperature can exhibit very high
values markedly exceeding $(10^4 - 10^5)$~V/W.

\section*{Acknowledgments}

The work was supported by the Japan Science and Technology Agency and the Japan Society for Promotion of Science, PIRE TeraNano Program, NSF, USA, and the Army Research Laboratory under ARL MSME Alliance, USA.

\end{document}